\documentclass[aps,prb,twocolumn,groupedaddress]{revtex4-2}

\usepackage{fancyhdr}
\usepackage{amsmath}
\usepackage{amssymb}
\usepackage{graphicx}
\usepackage{float}
\usepackage{subfigure}
\usepackage{dcolumn}
\usepackage{bm}
\usepackage[colorlinks,linkcolor=blue,citecolor=blue,urlcolor= blue]{hyperref}
\usepackage{times}
\usepackage{MnSymbol}
\usepackage[version=4]{mhchem} 
\usepackage{color}
\usepackage{siunitx}
\usepackage[title]{appendix}
\usepackage{verbatim}
\usepackage{comment}
\usepackage{csquotes}

\begin{document}


\title{Electric fields induced spin and/or valley polarization in Weiss oscillations of monolayer 1{\it T}$^{\prime}$-$\mathrm{MoS}_{2}$}
\author{Y. Li$^{1}$}
\author{W. Zeng$^{2}$}
\author{R. Shen$^{1,3,4}$}%
 \email{ E-mail: shen@nju.edu.cn}
\affiliation{%
$^{1}$National Laboratory of Solid State Microstructures and School of Physics, Nanjing University, Nanjing 210093, China\\
$^{2}$Department of Physics, Jiangsu University, Zhenjiang 212013, China\\
$^{3}$Collaborative Innovation Center of Advanced Microstructures, Nanjing University, Nanjing 210093, China\\
$^{4}$Jiangsu Key Laboratory of Quantum Information Science and Technology, Nanjing University, China 
}

\date{\today}

\begin{abstract}

Monolayer 1{\it T}$^{\prime}$-$\mathrm{MoS}_{2}$ exhibits spin- and valley-dependent massive tilted Dirac cones with two velocity correction terms in low-energy effective Hamiltonian. We theoretically investigate the longitudinal diffusive magneto-conductivity of monolayer 1{\it T}$^{\prime}$-$\mathrm{MoS}_{2}$ by using the linear response theory. It is shown that the Weiss oscillations are polarized in spin and valley degrees of freedom, under uniform electric fields and a weak one-dimensional spatially-periodic electrostatic potential modulation. The spin polarization, the valley polarization and the spin-valley polarization can be switched by flipping the external electric fields. The polarization is found not only in the amplitudes but also in the periods of the Weiss oscillations. It is found that the period polarization in Weiss oscillations originates from the polarized effective Fermi energies or the polarized Landau level spacing scales. In Weiss oscillations, polarization in amplitude does not imply the presence of polarization in period, whereas polarization in period is accompanied by polarization in amplitude. The superposition of polarization in amplitude and polarization in period enables the appearance of considerable polarization in Weiss oscillations under relatively weak external electric fields.

\end{abstract}

\maketitle\thispagestyle{fancy}

\section{\label{sec:level1}Introduction}

The discovery of graphene has triggered a great leap in the research on two-dimensional Dirac materials \cite{gra}. So far, numerous two-dimensional Dirac materials, for example, silicene \cite{25} and 8-{\it Pmmn} borophene \cite{20,21}, have been successfully fabricated and systematically studied both experimentally and theoretically. Silicene, with its controllable strong spin-orbit coupling gap, provides a great platform for spin transport research, while 8-{\it Pmmn} borophene, with its anisotropic tilted Dirac bands, paves the way for the manipulation of the valley degree of freedom. Over the past two decades, monolayer transition metal dichalcogenides (TMDCs) of the {\it T}$^{\prime}$ structure phase, which have the general chemical fomula $\mathrm{MX}_{2}$ with $\mathrm{M}$ ($\mathrm{W}$, $\mathrm{Mo}$) and $\mathrm{X}$ ($\mathrm{Te}$, $\mathrm{Se}$, $\mathrm{S}$) has attracted considerable attention \cite{1,2,3,4,5,6,7,8,9,10,11,12,13,14}. As a typical material of monolayer 1{\it T}$^{\prime}$-TMDCs, monolayer 1{\it T}$^{\prime}$-$\mathrm{MoS}_{2}$ \cite{1,15,16}, combining the aforementioned advantages of both silicene \cite{22,23,24} and 8-{\it Pmmn} borophene \cite{17,18,19}, has already led to various intriguing findings such as anisotropic longitudinal optical conductivities \cite{26}, anisotropic plasmon excitations and static screening effects \cite{27}, and electric field modulated valley- and spin-dependent Klein tunneling \cite{28}.

Magneto-transport measurements have always been appreciated for providing an efficient way to probe a two-dimensional fermionic system. In the presence of artificially created spatially-periodic modulation (electric, magnetic or both) with periods in the submicrometer range \cite{29,30,31,32,33,34}, it is known as Weiss oscillation \cite{29,30,35}, that the longitudinal diffusive conductivity oscillates with the inverse magnetic field periodically. Weiss oscillations have been investigated in two-dimensional electron gas (2DEG) \cite{43,44}, pristine graphene \cite{36,37}, bilayer graphene \cite{38}, $\alpha$-$\mathcal{T}_{3}$ lattices \cite{39}, phosphorene \cite{40}, graphene with surface acoustic waves \cite{41} and Galilean-invariant Dirac composite fermions \cite{42}. However, the polarization in amplitude of Weiss oscillations has always been the sole focus of researches \cite{45,46,47}, while the polarization of period remains relatively unexplored.

In this work, we theoretically investigate Weiss oscillations in monolayer 1{\it T}$^{\prime}$-$\mathrm{MoS}_{2}$ in the presence of uniform electric fields and a spatially-periodic electrostatic potential modulation in low temperature regime by using the linear response theory. In the presence/absence of uniform electric fields perpendicular to the material plane (vertical) and/or in-plane perpendicular to the current direction (transverse), there are four scenarios of Weiss oscillations. Firstly, the Weiss oscillation is unpolarized without external electric fields. Secondly, when a vertical electric field is applied, the curves of Weiss oscillations split into two branches. One branch represents Weiss oscillations for the spin-up electrons in the {\it K} valley or the spin-down electrons in the {\it K}$^{\prime}$ valley, while the other branch represents that for the spin-down electrons in the {\it K} valley or the spin-up electrons in the {\it K}$^{\prime}$ valley. This spin-valley polarization originates from the spin-valley-polarized effective Fermi energies and can be switched by flipping the vertical electric field. Thirdly, when a transverse electric field is applied, an effective tilting velocity is introduced and couples with the intrinsic tilting velocity, resulting in valley-polarized Landau level spacing scales and valley-polarized velocity correction factors. Consequently these valley-polarized factors give rise to valley polarization in Weiss  oscillations, which can be switched by flipping the transverse electric field. Fourthly, when both the vertical and the transverse electric fields are applied, apart from the valley polarization, the valley-polarized factors break down the spin-valley polarization, giving rise to the spin polarization. The valley polarization can only be switched by flipping the transverse electric field, while the spin polarization can be switched by either flipping the vertical or the transverse electric field. 

We found that the period polarization in Weiss oscillations originates from the polarized effective Fermi energies or the polarized Landau level spacing scales. In Weiss oscillations, polarization in amplitude does not imply the presence of polarization in period, whereas polarization in period is accompanied by polarization in amplitude. As long as either the vertical electric field or the transverse electric field is present, the polarization of Weiss oscillations in monolayer 1{\it T}$^\prime$-MoS$_{2}$ appears in both the amplitude and the period. The coexistence of amplitude polarization and period polarization leads to misalignment between the peaks and troughs of Weiss oscillation curves in non-equivalent channels. On the one hand, although the sign of the polarization observed solely from the perspective of amplitude or period is independent of magnetic field strength, the coupling of these two polarization types yields a polarization rate that can be either positive or negative with respect to different values of the magnetic field strength. This finding provides a potential approach for experimentally verifying the existence of period polarization in Weiss oscillations. On the other hand, the peak-trough misalignment ensures that the maximum polarization value remains largely unaffected with the electric field strength, enabling the appearance of considerable polarization in Weiss oscillations under relatively weak external electric fields.

The rest of the paper is organized as follows. In Sec.~\ref{sec:level2}, we introduce the low-energy effective Hamiltonian of monolayer 1{\it T}$^{\prime}$-$\mathrm{MoS}_{2}$ and obtain the Landau levels and eigenfunctions. The formalism calculating Weiss oscillations in the longitudinal diffusive magneto-conductivity is exhibited in Sec.~\ref{sec:level3}. In Sec.~\ref{sec:level4}, we present the results and make some discussions. Finally, we conclude in Sec.~\ref{sec:level5}.

\section{\label{sec:level2}Model Hamiltonian and Landau levels}
\label{sec:2}

\subsection{\label{sec:level2A}Model Hamiltonian}

We start with the low-energy $\bm{k} \cdot \bm{p}$ Hamiltonian of the monolayer 1{\it T}$^{\prime}$-$\mathrm{MoS}_{2}$ in the $x$-$y$ plane with a vertical electric field $E_{z}$ in the $z$ direction. The Hamiltonian in the $\bm{\sigma} \otimes \bm{s}$ space reads \cite{1,28}
\begin{equation}
\label{eq. H}
\mathcal{H} = \mathcal{H}_{\bm{k} \cdot \bm{p}} + \mathcal{H}_{E_{z}} + V,
\end{equation}where
\begin{equation}
\label{eq. Hkp}
\mathcal{H}_{\bm{k} \cdot \bm{p}} = \begin{pmatrix}
E_{p} & 0 & -iv_{1}\hbar q_{x} & v_{2}\hbar q_{y} \\
0 & E_{p} & v_{2}\hbar q_{y} & -iv_{1}\hbar q_{x} \\
iv_{1}\hbar q_{x} & v_{2}\hbar q_{y} & E_{d} & 0 \\
v_{2}\hbar q_{y} & iv_{1}\hbar q_{x} & 0 & E_{d}
\end{pmatrix} 
\end{equation}
and
\begin{equation}
\label{eq. HEz}
\mathcal{H}_{E_{z}} = w \alpha \Delta_{so}\begin{pmatrix}
0 & 0 & 1 & 0 \\
0 & 0 & 0 & 1 \\
1 & 0 & 0 & 0 \\
0 & 1 & 0 & 0
\end{pmatrix}.
\end{equation}
Here, $\bm{\sigma}$ stands for the Pauli matrix acting upon pseudospin space and $\bm{s}$ denotes the Pauli matrix acting upon real-spin space. The on-site energies of $p$ and $d$ orbitals are $E_{p}=\delta_{p} + \frac{\hbar^{2}q_{x}^{2}}{2m_{x}^{p}} + \frac{\hbar^{2}q_{y}^{2}}{2m_{y}^{p}}$ and $E_{d}=\delta_{d} + \frac{\hbar^{2}q_{x}^{2}}{2m_{x}^{d}} + \frac{\hbar^{2}q_{y}^{2}}{2m_{y}^{d}}$, respectively, where $q_{x,y}$ is the electron momentum, $\delta_{p} = 0.46 \mathrm{eV}$, $\delta_{d} = -0.20 \mathrm{eV}$, $m_{x}^{p} = -0.50 m_{0}$, $m_{y}^{p} = -0.16 m_{0}$, $m_{x}^{d} = 2.48 m_{0}$, $m_{y}^{d} = 0.37 m_{0}$, and $m_{0}$ is the free electron mass \cite{1}. $\delta_{p} > \delta_{d}$ at the $\Gamma$ point corresponds to the $p$-$d$ band inversion. The Fermi velocities along the $x$ and the $y$ directions are denoted as $v_{1} = 3.87 \times 10^{5} \mathrm{m/s}$ and $v_{2} = 0.46 \times 10^{5} \mathrm{m/s}$, respectively \cite{1}. Half of the fundamental spin-orbit coupling gap at the Dirac points is represented by $\Delta_{so}$, $w = \pm$ labels the direction of the vertical electric field $E_{z}$, and $\alpha$ is equal to $\left|E_{z}/E_{c}\right|$ with the critical electric field $E_{c} = 1.42 \mathrm{V/nm^{-1}}$ for topological phase transition \cite{1}. Potential $V$ can be adjusted by gate voltage or doping.

Under a proper unitary transformation
\begin{equation}
\label{eq. M}
M = \frac{1}{\sqrt{2}}\begin{pmatrix}
1 & 0 & 1 & 0 \\
-1 & 0 & 1 & 0 \\
0 & 1 & 0 & 1 \\
0 & -1 & 0 & 1
\end{pmatrix},
\end{equation} \\
the Hamiltonian $M^{-1}\mathcal{H}M$ is written in a block-diagonal form. After the linearization at two Dirac points $\Lambda = \pm(0,q_{0})$ with $q_{0} = 1.39 \mathrm{nm^{-1}}$, the diagonal term reads
\begin{equation}
\label{eq. Hzero}
\begin{split}
H_{s,\xi}(\bm{k}) &= \hbar k_{x}v_{1}\sigma_{y} - \hbar k_{y}(s v_{2}\sigma_{x} + \xi v_{-}\sigma_{0} + \xi v_{+}\sigma_{z}) \\
&+ \Delta_{so}(w \alpha - s\xi)\sigma_{x} + V\sigma_{0},
\end{split}
\end{equation}
where $s = \pm$ labels the spin index ($\uparrow$ or $\downarrow$), $\xi = \pm$ stands for the valley index ($K$ or $K^\prime$), and $k_{x,y}$ is the momentum measured from the Dirac points. In Eq. (\ref{eq. Hzero}), half of the fundamental spin-orbit coupling gap is given as $\Delta_{so} = \hbar v_{2} q_{0} = 42.1 \mathrm{meV}$, the first two terms correspond to the kinetic energies with two velocity correction terms $v_{-} = \frac{\hbar q_{0}}{2}(-\frac{1}{m_{y}^{p}} - \frac{1}{m_{y}^{d}}) = 2.86 \times 10^{5} \mathrm{m/s}$ and $v_{+} = \frac{\hbar q_{0}}{2}(-\frac{1}{m_{y}^{p}} + \frac{1}{m_{y}^{d}}) = 7.21 \times 10^{5} \mathrm{m/s}$ \cite{26}. The velocity correction term $v_{-}$, which is also known as the tilting velocity, describes the tilted nature of Dirac cones. The velocity correction term $v_{+}$ modifies the Fermi velocity in the $y$ direction and ensures that the Dirac cone will not be tipped over.

\subsection{\label{sec:level2B}Landau levels}

A vertical magnetic field $\bm{B} = B\bm{\hat{e}_{z}}$ can be introduced in the low-energy effective Hamiltonian in Eq. (\ref{eq. Hzero}) with the Landau gauge $\bm{A} = (0,Bx,0)$. After the Landau-Peierls substitution $\hbar\bm{k}\rightarrow\hbar\bm{k} + e\bm{A}$, one obtains 
\begin{equation}
\label{eq. HB}
\begin{split}
H_{s,\xi}(\bm{k}) &= \hbar k_{x} v_{1} \sigma_{y} - (\hbar k_{y} + e B x)(s v_{2} \sigma_{x} + \xi v_{-} \sigma_{0} + \xi v_{+} \sigma_{z}) \\
&+ \Delta_{so}(w \alpha - s \xi)\sigma_{x} + V \sigma_{0}.
\end{split}
\end{equation}
When the ratio of the transverse electric field $E_{x}$ to the vertical magnetic field $B$ is not too large ($v_{r} = E_{x}/{B} \ll 0.9v_{F}$ with $v_{F}$ labels the Fermi velocity in-plane perpendicular to the electric field), a uniform electric field along the $x$-direction can be treated as an effective tilt in $y$-direction deforming the Landau levels of two-dimensional Dirac materials, and the low energy effective Hamiltonian of two-dimensional Dirac materials is valid for calculation of Landau levels \cite{48}. The Hamiltonian with the transverse electric field $E_{x}$ reads
\begin{equation}
\label{eq. HBE}
\begin{split}
H_{s,\xi}(\bm{k}) &= r e E_{x} x \sigma_{0} + \hbar k_{x} v_{1} \sigma_{y} - (\hbar k_{y} + e B x)(s v_{2} \sigma_{x} \\
&+ \xi v_{-} \sigma_{0} + \xi v_{+} \sigma_{z}) + \Delta_{so}(w \alpha - s \xi)\sigma_{x} + V \sigma_{0},
\end{split}
\end{equation}
with $r = \pm$ labels the direction of $E_{x}$.
Noting that the commutator $[H,k_{y}] = 0$, we can write the wave function in the ansatz $\Psi(\bm{r}) = \frac{1}{\sqrt{L_{y}}}e^{ik_{y}y}\psi(x)$ and the Hamiltonian is simplified as
\begin{equation}
\label{eq. Hmatrix}
\begin{split}
&H = \\
&\begin{pmatrix}
-\xi \hbar(v_{+} + v_{-}^{\prime})[\sqrt{\frac{v_{1}}{v_{2}}} X + \frac{\xi \delta}{\sqrt{2}}] & -\hbar v_{c}[\partial_{X} + s X] \\
\hbar v_{c}[\partial_{X} - s X] & \xi \hbar(v_{+} - v_{-}^{\prime})[\sqrt{\frac{v_{1}}{v_{2}}} X + \frac{\xi \delta}{\sqrt{2}}]
\end{pmatrix} \\
&- r \epsilon_{r},
\end{split}
\end{equation}where $l = \sqrt{\hbar /eB}$ is the magnetic length, $\epsilon_{r} = \hbar v_{r} k_{y}$ is an additional energy correction term induced by the transverse electric field $E_{x}$, $v_{c} = \sqrt{v_{1}v_{2}}$ is the reduced velocity and $v_{-}^{\prime} = v_{-} - r \xi v_{r}$ is the corrected tilting velocity. The dimensionless position operator is defined as $X = \sqrt{\frac{v_{2}}{v_{1}}}\frac{x + x_{0}}{l}$ with $x_{0} = k_{y}l^{2} - \frac{\xi \delta l}{\sqrt{2}}$ where $\frac{\xi \delta l}{\sqrt{2}} = (w s \xi \alpha - 1)\frac{\Delta_{so}l^{2}}{\hbar v_{2}}$ is an additional shift in the cyclotron center induced by the spin-orbit coupling.

Following Refs. \cite{49,LY1}, we derive the eigenvalues and the eigenfunctions of the eigenproblem $H\psi(x) = \epsilon\psi(x)$ as:
\begin{equation}
\label{eq. epsilonky}
\begin{split}
&\epsilon_{s,\xi,\eta,n,k_{y}} = \frac{\sqrt{2} \hbar v_{c}}{l} \frac{1}{v_{2}^{2} + v_{+}^{2}} \Bigg[-\frac{1}{2} \delta v_{2} v_{-}^{\prime} \\
&+ \eta \sqrt{n \frac{(v_{2}^{2} + v_{+}^{2})(v_{2}^{2} + v_{+}^{2} - {v_{-}^{\prime}}^{2})^{3/2}}{v_{2}} + \frac{\delta^{2}}{4}v_{+}^{2}(v_{2}^{2} + v_{+}^{2} - {v_{-}^{\prime}}^{2})}\Bigg] \\
&- r \hbar v_{r} k_{y}, n = 0, 1, 2, \ldots,
\end{split}
\end{equation}

\begin{equation}
\label{eq. Psiup}
\begin{split}
&\Psi_{\uparrow,\xi,\eta,n,k_{y}}(\bm{r}) = \frac{1}{\sqrt{2 L_{y} l \theta}}e^{ik_{y}y} \\
&\begin{pmatrix}
-\xi(\frac{C_{+}}{C_{+} + C_{+}^{-1}})^{1/2}\phi_{n-1}(\tilde{X}) + \eta(\frac{C_{-}}{C_{-} + C_{-}^{-1}})^{1/2}\phi_{n}(\tilde{X}) \\
(\frac{C_{+}^{-1}}{C_{+} + C_{+}^{-1}})^{1/2}\phi_{n-1}(\tilde{X}) + \eta\xi(\frac{C_{-}^{-1}}{C_{-} + C_{-}^{-1}})^{1/2}\phi_{n}(\tilde{X})
\end{pmatrix},
\end{split}
\end{equation}

\begin{equation}
\label{eq. Psidown}
\begin{split}
&\Psi_{\downarrow,\xi,\eta,n}(\bm{r}) = \frac{1}{\sqrt{2 L_{y} l \theta}}e^{ik_{y}y} \\
&\begin{pmatrix}
-\xi(\frac{C_{+}^{-1}}{C_{+} + C_{+}^{-1}})^{1/2}\phi_{n-1}(\tilde{X}) + \eta(\frac{C_{-}^{-1}}{C_{-} + C_{-}^{-1}})^{1/2}\phi_{n}(\tilde{X}) \\
(\frac{C_{+}}{C_{+} + C_{+}^{-1}})^{1/2}\phi_{n-1}(\tilde{X}) + \eta\xi(\frac{C_{-}}{C_{-} + C_{-}^{-1}})^{1/2}\phi_{n}(\tilde{X})
\end{pmatrix},
\end{split}
\end{equation}with

\begin{equation}
\label{eq. Xtilde}
\tilde{X} = \frac{1}{l \theta}(x + \tilde{x}_{0}),
\end{equation}

\begin{equation}
\label{eq. xtilde}
\tilde{x}_{0} =k_{y} l^{2} - \frac{\xi \delta l}{\sqrt{2}} - \sqrt{2}\frac{\xi v_{-}^{\prime} \sqrt{\frac{v_{2}}{v_{1}}} \frac{(\epsilon_{n} + r \epsilon_{r}) l^{2}}{\sqrt{2} \hbar} - \frac{\xi \delta l}{2}(v_{+}^{2} - {v_{-}^{\prime}}^{2})}{v_{2}^{2} + v_{+}^{2} - {v_{-}^{\prime}}^{2}},
\end{equation}where $\eta = +1 (-1)$ is the band index, $n$ is the Landau-level index, $\theta = (\frac{v_{1}}{\sqrt{v_{2}^{2} + v_{+}^{2} - {v_{-}^{\prime}}^{2}}})^{1/2}$ is a parameter describing anisotropy, $\phi_{n}(\tilde{X})$ is the harmonic oscillator wave function with $\phi_{-1}(\tilde{X}) = 0$, and $C_{\pm} = \frac{v_{+} - v_{-}^{\prime}}{\sqrt{v_{2}^{2} + v_{+}^{2} - {v_{-}^{\prime}}^{2}} \mp v_{2}}$.

\section{\label{sec:level3}Formalism and calculation}
\label{sec:3}

In order to obtain the Weiss oscillation in monolayer 1{\it T}$^{\prime}$-$\mathrm{MoS}_{2}$ in low temperature regime, we assume a longitudinal current along the $y$ direction, the uniform electric field $E_{x}$ along the $x$ direction, the uniform electric field $E_{z}$ and the uniform magnetic field $B$ along the $z$ direction. We further introduce a weak spatially-periodic electrostatic potential modulation in the $x$ direction, which can be considered as a small perturbation. The conductivity in the $y$ direction is calculated by the Kubo formula \cite{50}. In order to show the conductivity correction due to the modulation, we calculate the longitudinal diffusive conductivity $\sigma_{yy}^{dif}$. Provided that the scattering processes are elastic or quasielastic, the diffusive conductivity is given by \cite{36,51,52}
\begin{equation}
\label{eq. Kubo}
\sigma_{yy}^{dif} = \frac{\beta e^{2}}{L_{x}L_{y}}\sum_{\zeta}f(\epsilon_{\zeta})[1 - f(\epsilon_{\zeta})]\tau_{\epsilon_{\zeta}}(v_{y}^{\zeta})^{2},
\end{equation}
where $L_{x}$, $L_{y}$ are the dimensions of the sample, $\zeta$ labels the quantum number of the electron eigenstate, $f(\epsilon_{\zeta}) = \{1 + \mathrm{exp}[\beta(\epsilon_{\zeta} - \epsilon_{F})]\}^{-1}$ is the Fermi-Dirac distribution function with the Fermi energy $\epsilon_{F}$ and the inverse temperature $\beta = (k_{B}T)^{-1}$, $\tau_{\epsilon_{\zeta}}$ denotes the energy-dependent collision time and $v_{y}^{\zeta}$ represents the group velocity.

The one-dimensional electrostatic potential modulation considered is described by the small perturbation Hamiltonian $H_{e}^{\prime} = V_{e}^{\prime}\mathrm{cos}(Kx)$ with $K = \frac{2\pi}{a_{0}}$, where $V_{e}^{\prime}$ is the modulation strength and $a_{0}$ is the period of spatial modulation. Using perturbation theory, we find the first-order energy correction as
\begin{equation}
\label{eq. Deltaepsilon}
\begin{split}
\Delta \epsilon_{e} &= \int_{-\infty}^{\infty} \,dx\int_{-L_{y}/2}^{L_{y}/2} \,dy\Psi^{\dagger}(\bm{r})H_{e}^{\prime}\Psi(\bm{r}) \\
&= \frac{V_{e}^{\prime}}{2}\{[F_{n-1}(u) + F_{n}(u)]\mathrm{cos}(K\tilde{x}_{0}) \\
&\quad- 2s\xi\eta\rho_{0}R_{n}(u)\mathrm{sin}(K\tilde{x}_{0})\},
\end{split}
\end{equation}
with
\begin{equation}
\label{eq. rho0}
\rho_{0} = \frac{C_{+}^{1/2}C_{-}^{1/2} - C_{+}^{-1/2}C_{-}^{-1/2}}{(C_{+} + C_{+}^{-1})^{1/2}(C_{-} + C_{-}^{-1})^{1/2}},
\end{equation}
\begin{equation}
\label{eq. Fnu}
F_{n}(u) = e^{-u/2}L_{n}(u),
\end{equation}
\begin{equation}
\label{eq. Rnu}
R_{n}(u) = \sqrt{\frac{2n}{u}}e^{-u/2}[L_{n-1}(u) - L_{n}(u)],
\end{equation}
where $L_{n}(u) = \sum\limits_{m = 0}^{n}\frac{(-1)^{m}n!u^{m}}{(m!)^{2}(n - m)!}$ is the Laguerre polynomial of order $n$ with $u = \frac{K^{2} l^{2} \theta^{2}}{2}$. Following Refs. \cite{36,37,47}, the diffusive conductivity can be simplified to an analytical form by using the higher Landau-level approximation
\begin{equation}
\label{eq. approLnu}
e^{-u/2}L_{n}(u) \rightarrow \frac{1}{\sqrt{\pi\sqrt{nu}}}\mathrm{cos}(2\sqrt{nu} - \frac{\pi}{4})
\end{equation}
and
\begin{equation}
\label{eq. approRnu}
R_{n}(u) \rightarrow \sqrt{2}\frac{1}{\sqrt{\pi\sqrt{nu}}}\mathrm{sin}(2\sqrt{nu} - \frac{\pi}{4}).
\end{equation}
We do not consider the case of a very high magnetic field, where the SdH oscillations dominate the conductivity, so that the employment of the asymptotic expression for the Laguerre polynomials is a good approximation \cite{36}.

The energy correction can be written as
\begin{equation}
\label{eq. Deltaepsilon2}
\begin{split}
\Delta\epsilon_{e} = V_{e}^{\prime}\frac{1}{\sqrt{\pi\sqrt{nu}}}&[\mathrm{cos}(2\sqrt{nu} - \frac{\pi}{4})\mathrm{cos}(K\tilde{x}_{0}) \\
&- \sqrt{2}s\xi\eta\rho_{0}\mathrm{sin}(2\sqrt{nu} - \frac{\pi}{4})\mathrm{sin}(K\tilde{x}_{0})],
\end{split}
\end{equation}
which finally leads to the nonzero drift velocity
\begin{equation}
\label{eq. vy}
\begin{split}
v_{y} = \frac{1}{\hbar}\frac{\partial\Delta\epsilon_{e}}{\partial k_{y}} = -\frac{2V_{e}^{\prime}}{\hbar K {\theta}^{2}}&u\frac{1}{\sqrt{\pi\sqrt{nu}}}[\mathrm{cos}(2\sqrt{nu} - \frac{\pi}{4})\mathrm{sin}(K\tilde{x}_{0}) \\
&+ \sqrt{2}s\xi\eta\rho_{0}\mathrm{sin}(2\sqrt{nu} - \frac{\pi}{4})\mathrm{cos}(K\tilde{x}_{0})].
\end{split}
\end{equation}

The summation in Eq. (\ref{eq. Kubo}) can be represented by
\begin{equation}
\label{eq. sumzeta}
\sum_{\zeta} = \frac{L_{y}}{2\pi}\int_{-L_{x}/2l^2}^{L_{x}/2l^2}dk_{y}\sum_{s,\xi,\eta,n}.
\end{equation}
With the help of Eq. (\ref{eq. vy}), one can obtain the final expression for the diffusive dc conductivity
\begin{equation}
\label{eq. sigmayydif}
\sigma_{yy}^{dif} = \frac{e^{2}}{h}W_{e}\Phi,
\end{equation}
where
\begin{equation}
\label{eq. We}
W_{e} = \frac{V_{e}^{\prime 2}\tau_{0}\beta}{\hbar}
\end{equation}
is the dimensionless strength of the electric modulation and
\begin{equation}
\label{eq. Phi}
\begin{split}
\Phi &= \frac{\sqrt{u}}{\pi {\theta}^{2}}\sum_{s,\xi,\eta,n}\frac{g}{(1 + g)^{2}}\frac{1}{\sqrt{n}}[\mathrm{cos}^{2}(2\sqrt{nu} - \frac{\pi}{4}) \\
&+ 2\rho_{0}^{2}\mathrm{sin}^{2}(2\sqrt{nu} - \frac{\pi}{4})] \\
&= \sum_{s,\xi} \Phi_{s,\xi}
\end{split}
\end{equation}
is the dimensionless conductivity with the exponential function $g = \mathrm{exp}\{\beta[\epsilon_{s,\xi,\eta,n} - \epsilon_{F}]\}$. Here, we assume that the collisional time $\tau_{\epsilon_{\zeta}}$ is a constant $\tau_{0}$, which is a good approximation when the temperature is sufficiently low \cite{36,37,46}. The dimensionless longitudinal diffusive magneto-conductivity in a specific spin channel and a specific valley channel reads
\begin{equation}
\label{eq. Phisxi}
\begin{split}
\Phi_{s,\xi} = \frac{\sqrt{u}}{\pi {\theta}^{2}}\sum_{\eta,n}\frac{g}{(1 + g)^{2}}\frac{1}{\sqrt{n}}&[\mathrm{cos}^{2}(2\sqrt{nu} - \frac{\pi}{4}) \\
&+ 2\rho_{0}^{2}\mathrm{sin}^{2}(2\sqrt{nu} - \frac{\pi}{4})].
\end{split}
\end{equation}

We set the sample size $L_{x} \times L_{y}$ from $10 a_{0} \times 10 a_{0}$ to  $100 a_{0} \times 100 a_{0}$ with the modulation period $a_{0} = 350 \mathrm{nm}$ and apply a weak transverse electric field $E_{x} \leq 0.2 B\sqrt{v_{2}^{2} + v_{+}^{2}}$, so that the the polarization-independent $k_{y}$ energy correction term in Eq. (\ref{eq. epsilonky}) can be neglected, and we can obtain a simpler analytical expression for the diffusive conductivity. The simplified Landau level reads
\begin{equation}
	\label{eq. epsilon}
	\begin{split}
		&\epsilon_{s,\xi,\eta,n} = \frac{\sqrt{2} \hbar v_{c}}{l} \frac{1}{v_{2}^{2} + v_{+}^{2}} \Bigg[-\frac{1}{2} \delta v_{2} v_{-}^{\prime} \\
		&+ \eta \sqrt{n \frac{(v_{2}^{2} + v_{+}^{2})(v_{2}^{2} + v_{+}^{2} - {v_{-}^{\prime}}^{2})^{3/2}}{v_{2}} + \frac{\delta^{2}}{4}v_{+}^{2}(v_{2}^{2} + v_{+}^{2} - {v_{-}^{\prime}}^{2})}\Bigg].
	\end{split}
\end{equation}
Here, we explicitly denote that the Landau level depends on $s$, $\xi$, $\eta$ and $n$ for clarity.

The Landau level presented in Eq. (\ref{eq. epsilon}) can be split into two parts. One is a constant term proportional to $\delta$, the other is relevant to $\sqrt{n}$. The constant term is independent on the Landau index $n$, and can be absorbed into $\epsilon_{F}$ leading to an effective Fermi energy $\epsilon_{F}^{\prime} = \epsilon_{F} + \frac{\sqrt{2} \hbar v_{c}}{l} \frac{1}{v_{2}^{2} + v_{+}^{2}} \frac{1}{2} \delta v_{2} v_{-}^{\prime}$. Similarly, by some rough approximation, the $n$-independent gap term in the square root in Eq. (\ref{eq. epsilon}) can also be absorbed into the effective Fermi energy as $\epsilon_{F}^{\prime \prime} = \sqrt{{\epsilon_{F}^{\prime}}^{2} - (\frac{\sqrt{2} \hbar v_{c}}{l} \frac{1}{v_{2}^{2} + v_{+}^{2}})^{2} \frac{\delta^{2}}{4}v_{+}^{2}(v_{2}^{2} + v_{+}^{2} - {v_{-}^{\prime}}^{2})}$ since only the Landau levels near the Fermi energy determines the conductivity. 

By introducing the effective Fermi energy, the dimensionless conductivity $\Phi_{s,\xi}$ in Eq. (\ref{eq. Phisxi}) can be simplified as
\begin{equation}
	\label{eq. Phisxi2}
	\begin{split}
		\Phi_{s,\xi} &= \frac{T}{4 \pi^{2} {\theta}^{2} T_{D}} \mathrm{cos}^{2}(\frac{\pi \hbar v_{h}}{\epsilon_{F}^{\prime \prime} a_{0}}) [(1 + 2 \rho_{0}^{2}) \\
		&+ (1 - 2 \rho_{0}^{2}) \mathrm{cos}(\beta \epsilon_{F}^{\prime \prime} \frac{T}{T_{D}} - \frac{\pi}{2}) \frac{T/T_{D}}{\mathrm{sinh}(T/T_{D})}]
	\end{split}
\end{equation}
with the effective Fermi energy $\epsilon_{F}^{\prime \prime}$, the characteristic temperature $T_{D} = \frac{\hbar v_{h} B}{4 \pi^{2} a_{0} k_{B} B_{0}}$, and the characteristic velocity $v_{h} = \sqrt{\frac{v_{1}(v_{2}^{2} + v_{+}^{2} - {v_{-}^{\prime}}^{2})^{3/2}}{v_{2}^{2} + v_{+}^{2}}}$, which scales the Landau level spacing.

\section{\label{sec:level4}Results and discussions}
\label{sec:4}

In the presence/absence of the uniform vertical and/or transverse electric fields, there are four scenarios of Weiss oscillations in monolayer 1{\it T}$^{\prime}$-$\mathrm{MoS}_{2}$. For each scenario, we consider the valley and/or spin polarization of the Weiss oscillation under the vertical and/or transverse electric fields with different strengths and directions. In this section, we plot the numerical curves for Weiss oscillations by using Eq. (\ref{eq. Phisxi}), and conduct analytical discussions by using Eq. (\ref{eq. Phisxi2}).

\subsection{Presence of the uniform vertical electric field}
\label{subsec:Presence of the uniform vertical electric field}

\begin{figure}[h]
\centering
\includegraphics[width=0.9\linewidth]{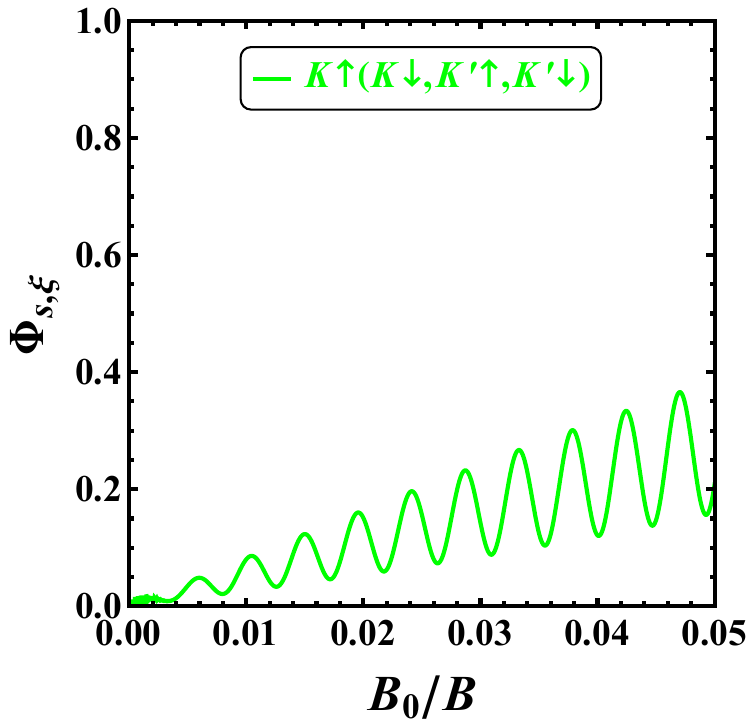}
\caption{\label{fig:1}The diffusive conductivities in different spin and valley channels versus the inverse magnetic field in the absence of both electric fields. The modulation period is $a_{0} = 350 \mathrm{nm}$, the system size is $L_{x} \times L_{y} = 10 a_{0} \times 10 a_{0}$, the temperature is $T = 3 \mathrm{K}$, and the Fermi energy is taken as 0.17 eV. $B_{0} = \frac{\hbar}{ea_{0}^{2}}$ is the characteristic magnetic field.}
\end{figure}

As can be seen in Fig.\ \ref{fig:1}, the Weiss oscillation is unpolarized without external electric fields.

\begin{figure}[h]
\centering
\subfigbottomskip=2pt
\subfigcapskip=-5pt
\subfigure[$w = 1, \alpha = 0.2$]{
\label{fig:2a}
\includegraphics[width=0.45\linewidth]{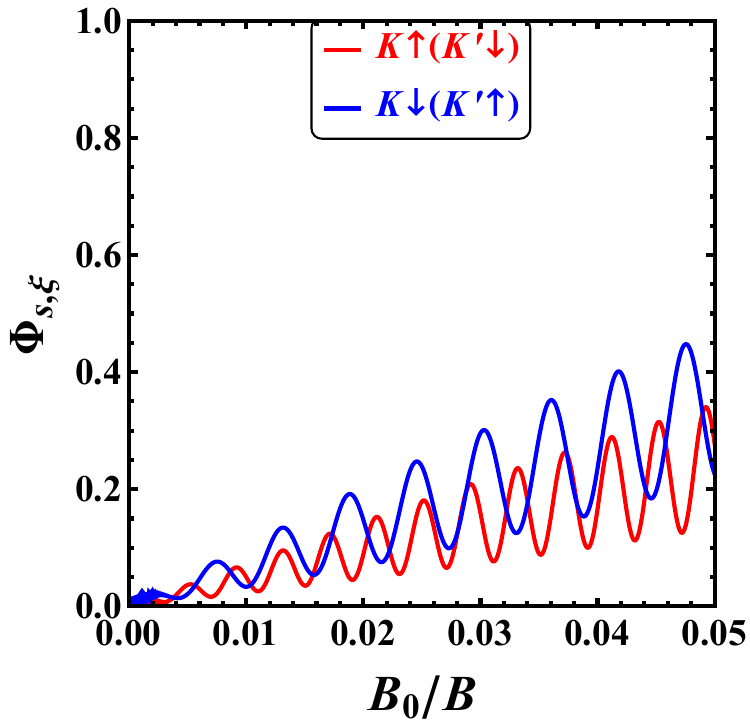}
}
\subfigure[$w = 1, \alpha = 0.4$]{
\label{fig:2b}
\includegraphics[width=0.45\linewidth]{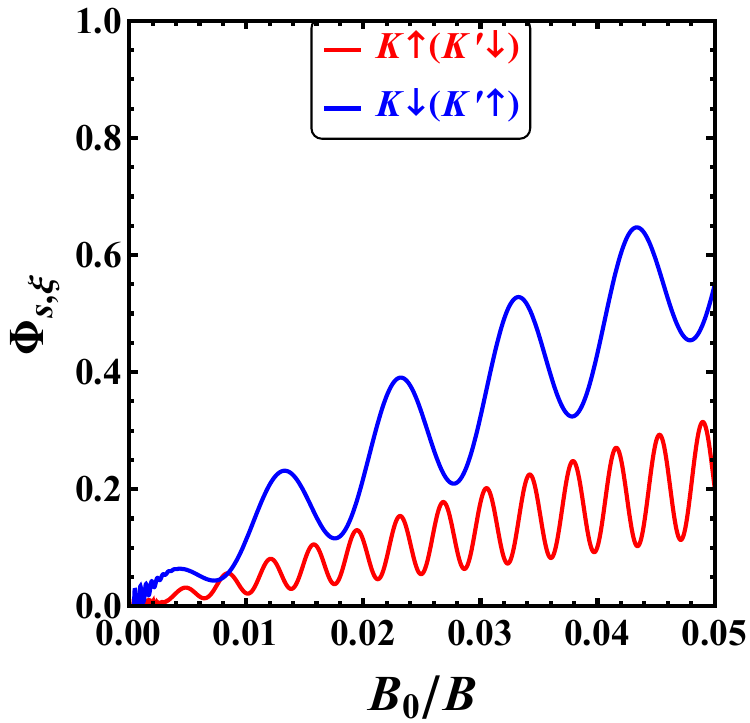}
}
 \\
\subfigure[$w = 1, \alpha = 0.2$]{
\label{fig:2c}
\includegraphics[width=0.45\linewidth]{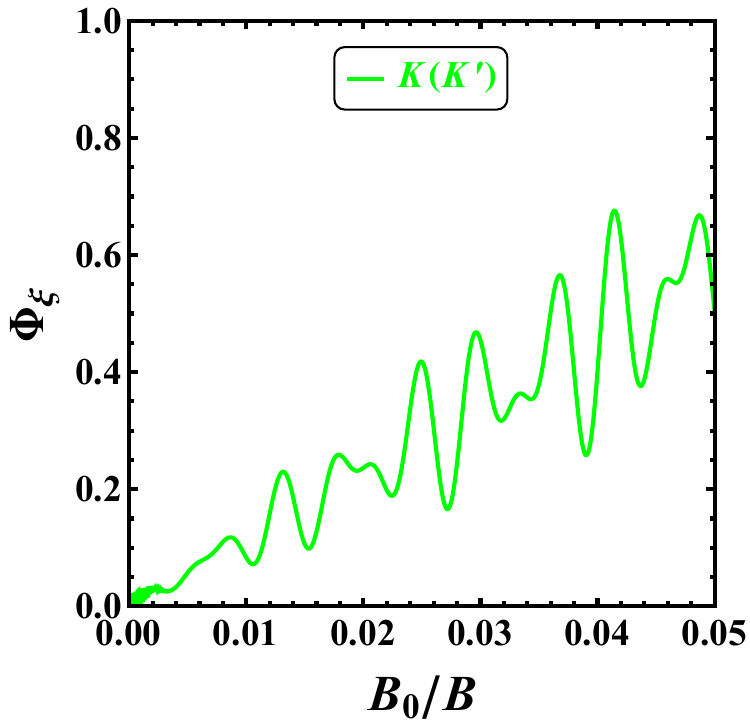}
}
\subfigure[$w = 1, \alpha = 0.2$]{
\label{fig:2d}
\includegraphics[width=0.45\linewidth]{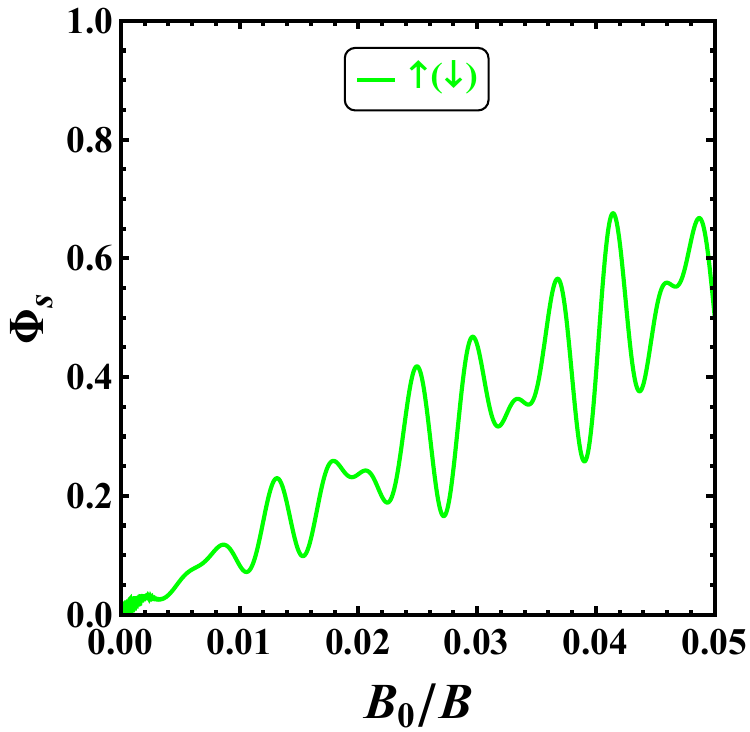}
}
\caption{\label{fig:2}The diffusive conductivities versus the inverse magnetic field in the presence of the uniform vertical electric field.}

\end{figure}

It is shown in Fig.\ \ref{fig:2a} and Fig.\ \ref{fig:2b} that the curves of the dimensionless conductivity $\Phi_{s,\xi}$ split into two branches in the presence of the uniform vertical electric field, resulting in the spin-valley polarization in the Weiss oscillation. The red curves in Fig.\ \ref{fig:2a} and Fig.\ \ref{fig:2b} represent Weiss oscillations for the spin-up electrons in the {\it K} valley or the spin-down electrons in the {\it K}$^{\prime}$ valley, while the blue curves represent that for the spin-down electrons in the {\it K} valley or the spin-up electrons in the {\it K}$^{\prime}$ valley. Comparing Fig.\ \ref{fig:2b} with Fig.\ \ref{fig:2a}, one finds that the amplitude and the period of the Weiss oscillations in the {\it K}$\uparrow$ and {\it K}$^{\prime}$$\downarrow$ channel decrease while these in the {\it K}$\downarrow$ and {\it K}$^{\prime}$$\uparrow$ channel increase with finite $E_{z}$. This spin-valley polarization is found not only in the amplitudes but also in the periods of the Weiss oscillations. In the presence of $E_{z}$, the Rashba spin-orbit coupling is introduced into the system and couples with the intrinsic one, leading to spin-valley-polarized gaps and consequently spin-valley-polarized effective Fermi energies, making contributions to the spin-valley polarization in Weiss oscillations.

From Eq. (\ref{eq. Phisxi2}), we naturally draw the conclusion that the period polarization in Weiss oscillations originates only from the polarized effective Fermi energies or the polarized Landau level spacing scales. Furthermore, it is also obvious that the polarized effective Fermi energies or the polarized Landau level spacing scales necessarily leads to the amplitude polarization in Weiss oscillations. Therefore, in Weiss oscillations, polarization in amplitude, for example, in borophene \cite{47}, does not imply the presence of polarization in period, whereas polarization in period is accompanied by polarization in amplitude as in monolayer 1{\it T}$^{\prime}$-$\mathrm{MoS}_{2}$. 

However, this spin-valley polarization in Weiss oscillations is not a completely polarized state. As shown in Fig.\ \ref{fig:2c} and Fig.\ \ref{fig:2d}, when considering a single valley channel or a single spin channel, the Weiss oscillation is unpolarized. In addition, although curves of dimensionless conductivities in different spin and valley channels agree with Weiss oscillations of general two-dimensional material (standard or Dirac), oscillating with a certain period and a linearly increasing amplitude \cite{44,36}, the total Weiss oscillation curve of monolayer 1{\it T}$^{\prime}$-$\mathrm{MoS}_{2}$ become less regular due to misalignments of peaks and troughs of Weiss oscillation curves from non-equivalent channels, which is a result of the superposition of the polarization in amplitude and the polarization in period.

\begin{figure}[h]
\centering
\subfigbottomskip=2pt
\subfigcapskip=-5pt
\subfigure[$\alpha = 0.2$]{
\label{fig:3a}
\includegraphics[width=0.45\linewidth]{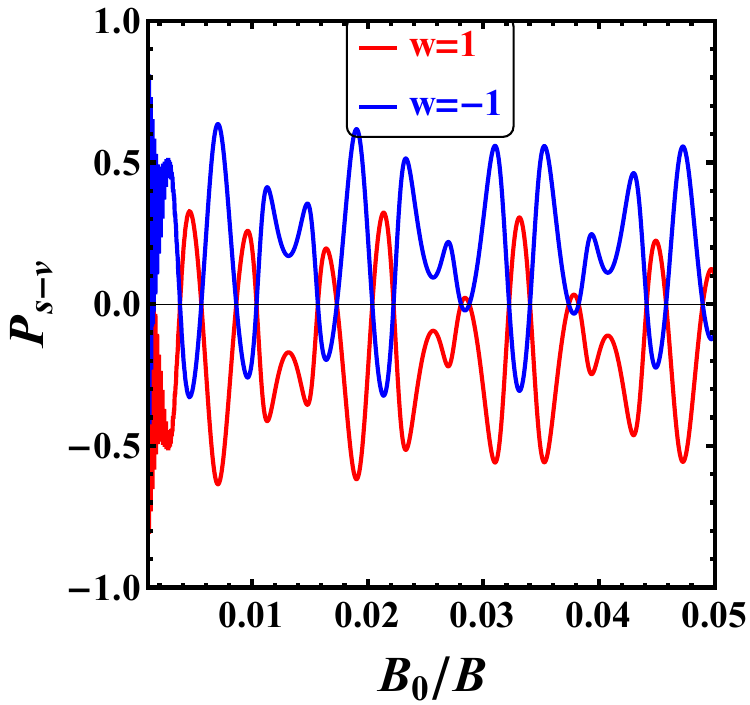}
}
\subfigure[$w = 1$]{
\label{fig:3b}
\includegraphics[width=0.45\linewidth]{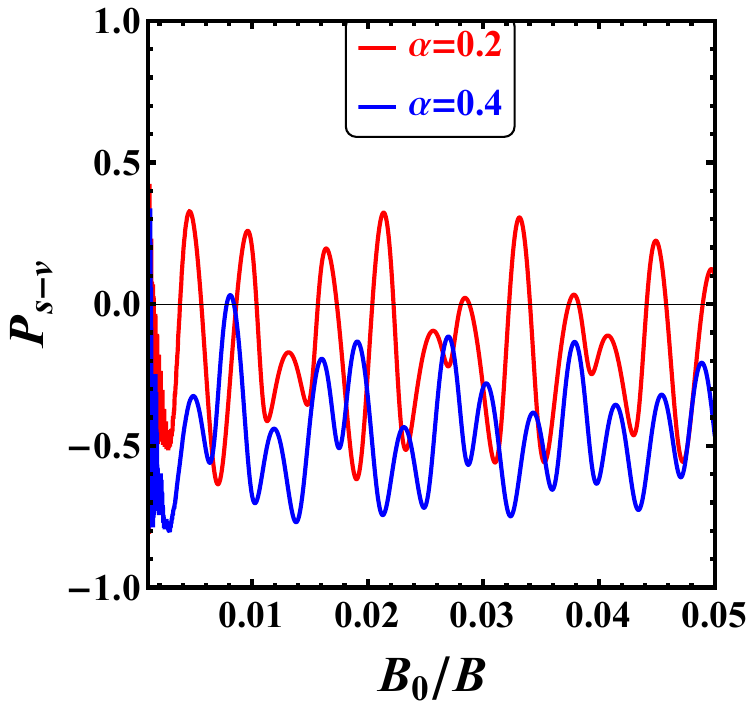}
}
\caption{\label{fig:3}The polarization rates versus the inverse magnetic field in the presence of the uniform vertical electric field.}
\end{figure}

Defining the spin-valley polarization rate as
\begin{equation}
\label{eq. psv}
P_{s-v} = \frac{\sigma_{yy,s \xi = 1}^{dif} - \sigma_{yy,s \xi = -1}^{dif}}{\sigma_{yy,s \xi = 1}^{dif} + \sigma_{yy,s \xi = -1}^{dif}},
\end{equation}
and plotting the spin-valley polarization rates under a uniform vertical electric field with opposite directions in Fig.\ \ref{fig:3a}, one finds that this spin-valley polarization rate can be flipped by flipping the direction of $E_{z}$. This can be understood analytically since the direction label $w$ and the spin index $s$ always appear in the form of a product in the expression of $\Phi_{s,\xi}$ in Eq. (\ref{eq. Phisxi}). By reversing the direction of the vertical electric field, two unequal spin-orbit coupling gaps are exchanged, thereby exchanging the effective Fermi surfaces and reversing the spin-valley-polarized current.

The curves of spin-valley polarization rates under uniform vertical electric fields with the same direction but different strengths are plotted in Fig.\ \ref{fig:3b}. It is shown that although the sign of the polarization rate observed solely from the perspective of amplitude or period is independent of magnetic field strength, the coupling of these two polarization types yields a polarization rate that can be either positive or negative with respect to different values of the magnetic field strength. In addition, we find it that the spin-valley polarization rate is actually not necessarily positively correlated with the value of the vertical electric field for a given magnetic field. This is due to misalignments of peaks and troughs of polarization rate curves from non-equivalent channels, which is a result of the superposition of the polarization in amplitude and the polarization in period. Therefore, we might be able to obtain considerable polarization in Weiss oscillations under relatively weak external electric fields.

\subsection{Presence of the uniform transverse electric field}
\label{subsec:Presence of the uniform transverse electric field}

\begin{figure}[h]
\centering
\subfigbottomskip=2pt
\subfigcapskip=-5pt
\subfigure[$r = 1, t_{r} = 0.1$]{
\label{fig:4a}
\includegraphics[width=0.45\linewidth]{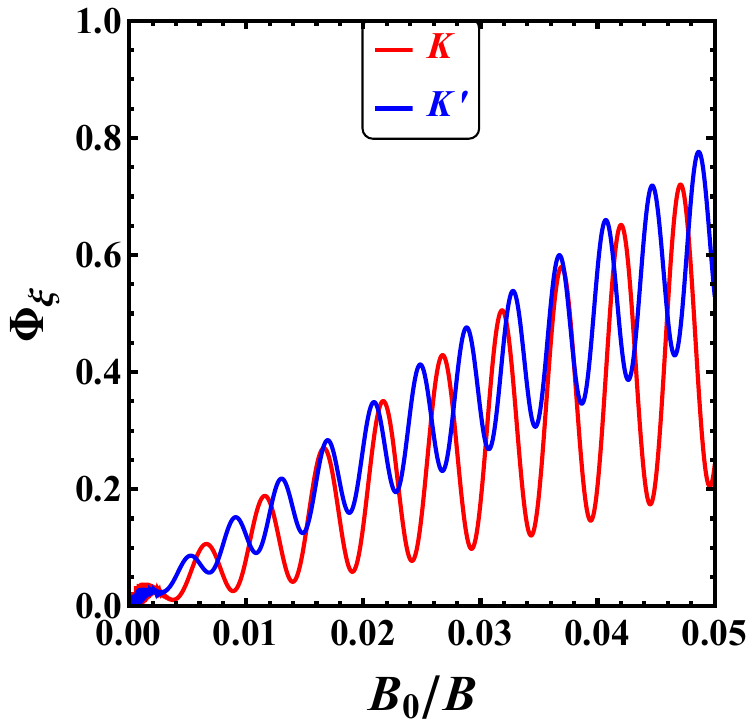}
}
\subfigure[$r = 1, t_{r} = 0.1$]{
\label{fig:4b}
\includegraphics[width=0.45\linewidth]{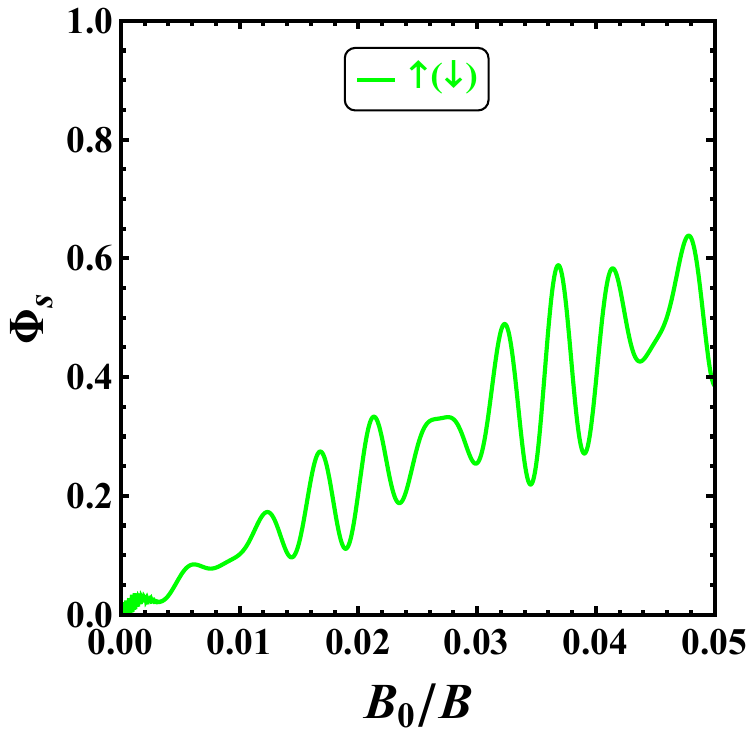}
}
\caption{\label{fig:4}The diffusive conductivities versus the inverse magnetic field in the presence of the uniform transverse electric field.}
\end{figure}

In the presence of the uniform transverse electric field, an effective tilting velocity $v_{r} = \frac{E_{x}}{B}$ was introduced. Unlike the intrinsic tilting of the monolayer 1{\it T}$^{\prime}$-$\mathrm{MoS}_{2}$, where two Dirac cones at different valleys tilt towards each other, the transverse electric field causes the non-equivalent Dirac cones tilt in the same direction. Coupled by these two tilting velocities, the corrected tilting velocities $v_{-}^{\prime} = v_{-} - r \xi v_{r}$ generate unequal drift velocities along the $y$ direction at two Dirac points as shown in Eq. (\ref{eq. vy}), leading to valley-polarized Weiss oscillations. Here, we define the effective tilting parameter $t_{r} = \frac{v_{r}}{\sqrt{v_{2}^{2} + v_{+}^{2}}}$, and plot the diffusive conductivities in specific valleys versus the inverse magnetic field in Fig.\ \ref{fig:4a}. Comparing Fig.\ \ref{fig:4a} with Fig.\ \ref{fig:1}, the curves of the dimensionless conductivity $\Phi_{s,\xi}$ split into two branches when a nonzero $E_{x}$ was applied, where the period of the Weiss oscillation in the {\it K} valley increases while that in the {\it K}$^{\prime}$ valley decreases. It is also shown in Fig.\ \ref{fig:4a} that the amplitude of the Weiss oscillation in the {\it K} valley is smaller than that in the {\it K}$^{\prime}$ valley. It is shown in Fig.\ \ref{fig:4b} that the Weiss oscillation is spin-unpolarized in the presence of the uniform transverse electric field. Similar to the scenario applying $E_{z}$, the curve of the total Weiss oscillation in the presence of $E_{x}$ is also irregular due to misalignments of peaks and troughs of polarization rate curves from different valley channels, which is a result of the superposition of the polarization in amplitude and the polarization in period.

\begin{figure}[h]
\centering
\includegraphics[width=0.9\linewidth]{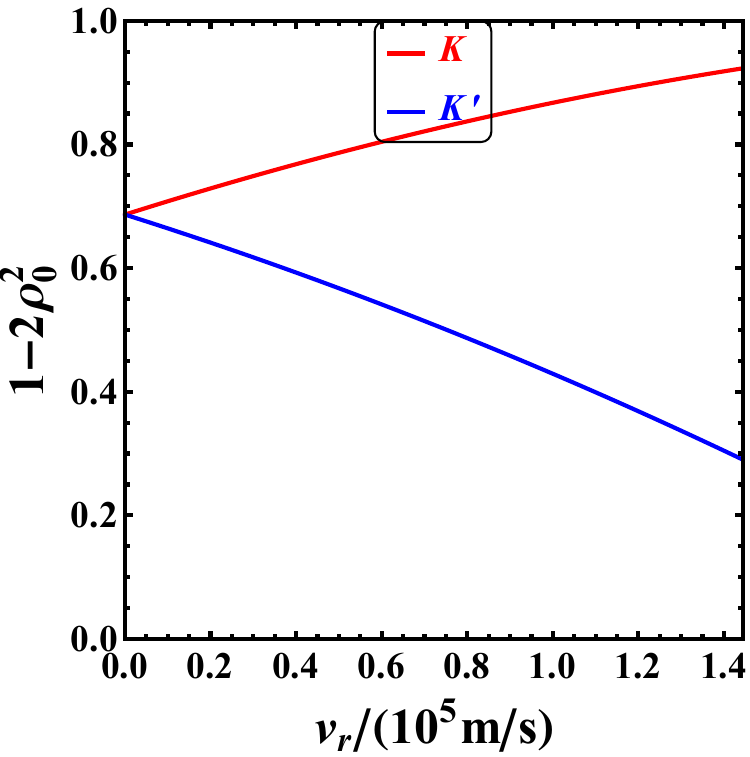}
\caption{\label{fig:5}The velocity correction factor $1 - 2 \rho_{0}^{2}$ in different valleys versus the effective tilting velocity $v_{r}$.}
\end{figure}

This polarization behavior can be understood analytically by comparing Eq. (\ref{eq. Phisxi2}) with the results in graphene \cite{36}, a typical two-dimensional Dirac material. Firstly, a smaller corrected tilting velocity brings in a larger Landau level spacing scale, resulting in a larger period and a smaller amplitude of the Weiss oscillation. Secondly, a smaller corrected tilting velocity corresponds to a larger anisotropy parameter $\theta$, leads to a smaller amplitude modulation factor $1 / {\theta}^{2}$ and a smaller amplitude of the Weiss oscillation. Thirdly, a smaller corrected tilting velocity corresponds to a smaller amplitude modulation factor $\mathrm{cos}^{2}(\frac{\pi \hbar v_{h}}{\epsilon_{F}^{\prime \prime} a_{0}})$ for the parameters we have selected here. Fourthly, as plotted in Fig.\ \ref{fig:5}, a smaller $v_{-}^{\prime}$ means a larger velocity correction factor $1 - 2 \rho_{0}^{2}$ in Eq. (\ref{eq. Phisxi2}), leading to a larger amplitude. As a result, a smaller corrected tilting velocity leads to a definitely smaller oscillation period due to the larger landau level spacing scale but a slightly change in the amplitude determined by the competition between the larger Landau level spacing scale, the larger anisotropy parameter $\theta$, the smaller amplitude modulation factor $\mathrm{cos}^{2}(\frac{\pi \hbar v_{h}}{\epsilon_{F}^{\prime \prime} a_{0}})$ and the larger velocity correction factor $1 - 2 \rho_{0}^{2}$.

\begin{figure}[h]
\centering
\subfigbottomskip=2pt
\subfigcapskip=-5pt
\subfigure[$t_{r} = 0.1$]{
\label{fig:6a}
\includegraphics[width=0.45\linewidth]{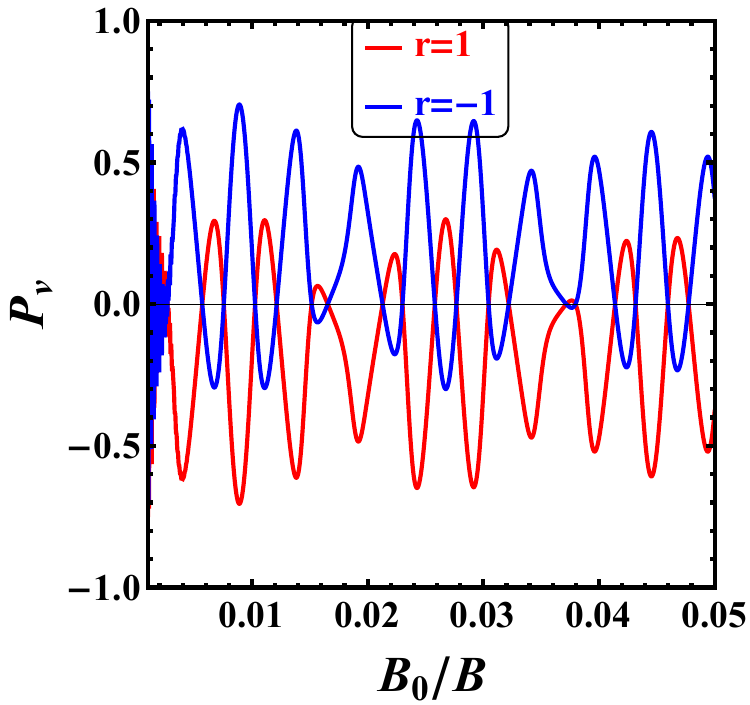}
}
\subfigure[$r = 1$]{
\label{fig:6b}
\includegraphics[width=0.45\linewidth]{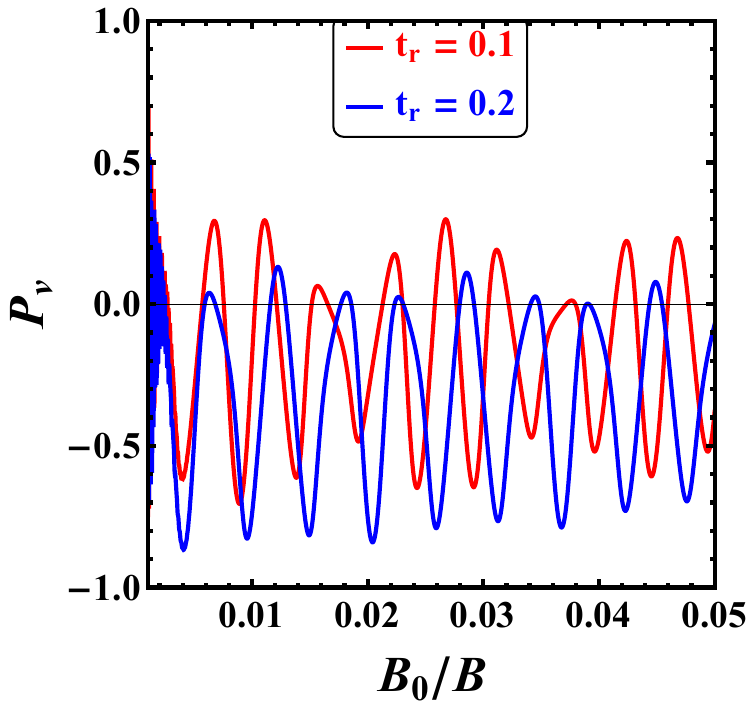}
}
\caption{\label{fig:6}The polarization rates versus the inverse magnetic field in the presence of the uniform transverse electric field.}
\end{figure}

Defining the valley polarization rate as
\begin{equation}
\label{eq. pv}
P_{v} = \frac{\sigma_{yy,\xi = 1}^{dif} - \sigma_{yy,\xi = -1}^{dif}}{\sigma_{yy,\xi = 1}^{dif} + \sigma_{yy,\xi = -1}^{dif}},
\end{equation}
and plotting the valley polarization rates under a uniform transverse electric field with opposite directions in Fig.\ \ref{fig:6a}, one finds that this valley polarization can be flipped by flipping the direction of $E_{x}$. This result is obvious because the direction label $r$ and the spin index $\xi$ always appear in the form of a product in the expression of $\Phi_{s,\xi}$ in Eq. (\ref{eq. Phisxi}). Coupling of the intrinsic tilting and the $E_{x}$-introduced effective tilting ensures that one can select the valley conductivity by switching the direction of the uniform transverse electric field.

The curves of valley polarization rates under uniform transverse electric fields with the same direction but different strengths are plotted in Fig.\ \ref{fig:6b}. Similar to the scenario applying the uniform vertical electric field, due to the superposition between the polarization in amplitude and the polarization in period, the misalignments of peaks and troughs in Weiss oscillation curves ensure that we can obtain a sizable spin/valley polarization with rather weak external electric fields.

\subsection{Presence of both the uniform vertical and transverse electric fields}
\label{subsec:Presence of both the uniform vertical and transverse electric fields}

\begin{figure}[h]
\centering
\subfigbottomskip=2pt
\subfigcapskip=-5pt
\subfigure[$w = 1, r = 1, \alpha = 0.2, t_{r} = 0.1$]{
\label{fig:7a}
\includegraphics[width=0.45\linewidth]{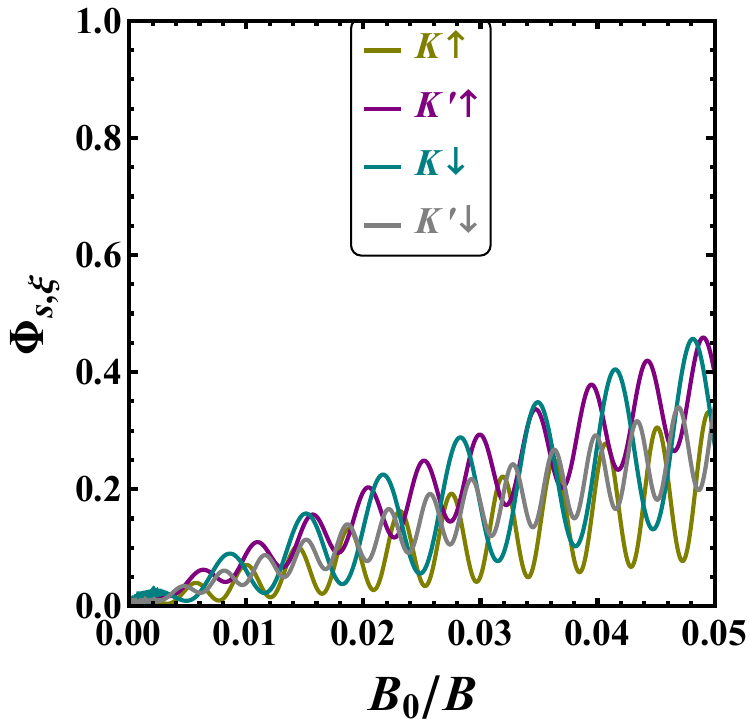}
}
\subfigure[$w = 1, r = 1, \alpha = 0.2, t_{r} = 0.1$]{
\label{fig:7b}
\includegraphics[width=0.45\linewidth]{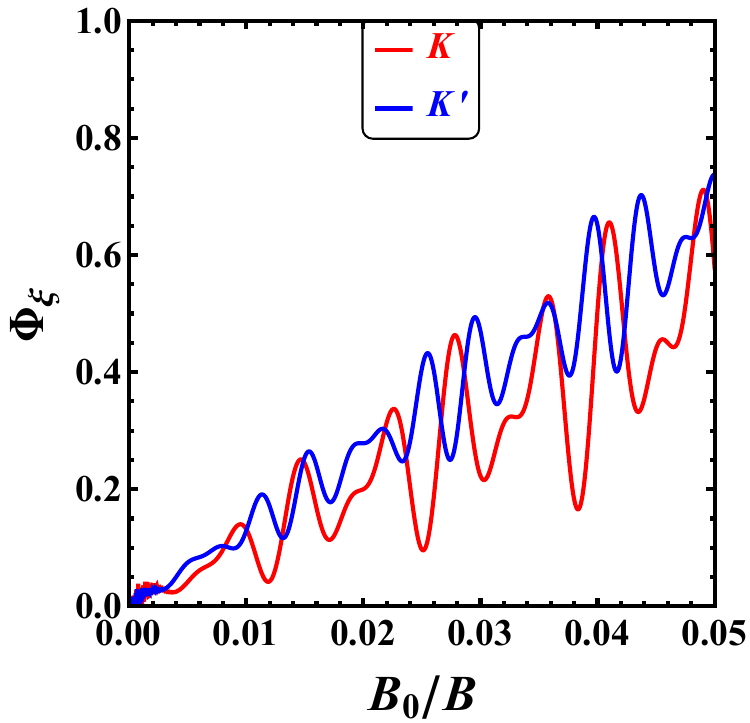}
}
 \\
\subfigure[$w = 1, r = 1, \alpha = 0.2, t_{r} = 0.1$]{
\label{fig:7c}
\includegraphics[width=0.45\linewidth]{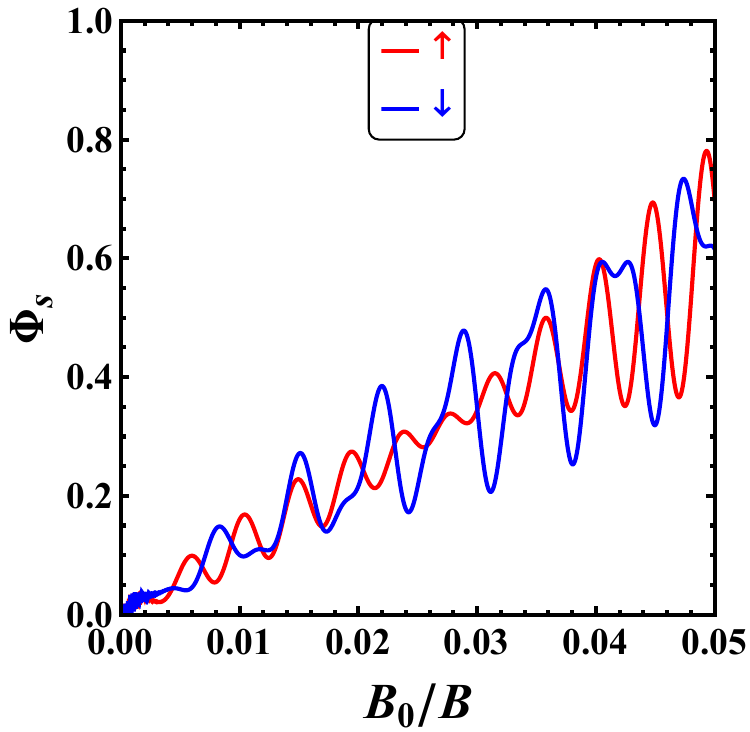}
}
\subfigure[$w = 1, r = 1, \alpha = 0.2, t_{r} = 0.1$]{
\label{fig:7d}
\includegraphics[width=0.45\linewidth]{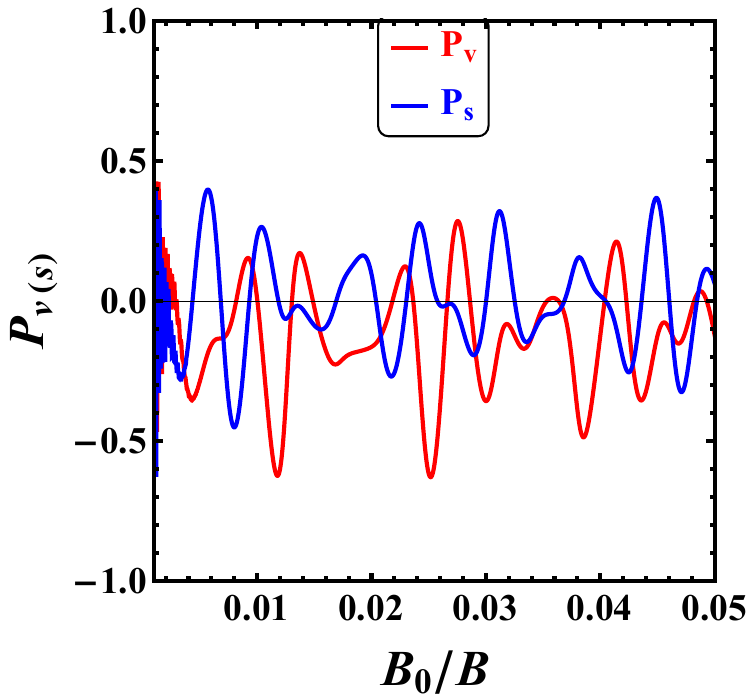}
}
\caption{\label{fig:7}The diffusive conductivities or polarization rates versus the inverse magnetic field in the presence of both the uniform vertical and transverse electric fields.}
	
\end{figure}

It is shown in Fig.\ \ref{fig:7a} that, when both the uniform vertical and transverse electric fields are applied, the curves of Weiss oscillations in monolayer 1{\it T}$^{\prime}$-$\mathrm{MoS}_{2}$ split into four branches, leading to the valley polarization plotted in Fig.\ \ref{fig:7b} and the spin polarization plotted in Fig.\ \ref{fig:7c}. By defining the spin polarization rate as
\begin{equation}
\label{eq. ps}
P_{s} = \frac{\sigma_{yy,s = 1}^{dif} - \sigma_{yy,s = -1}^{dif}}{\sigma_{yy,s = 1}^{dif} + \sigma_{yy,s = -1}^{dif}},
\end{equation}
we plot the valley polarization rate $P_{v}$ and the spin polarization rate $P_{s}$ in Fig.\ \ref{fig:7d}. Similar to scenarios where only $E_{z}$ or only $E_{x}$ is applied, both the amplitude and the period take part in the polarizations of Weiss oscillations.

In the presence of $E_{z}$, the Rashba spin-orbit coupling is introduced into the system and couples with the intrinsic one, leading to spin-valley-polarized gaps and consequently spin-valley-polarized effective Fermi energies, making contributions to the spin-valley polarization in Weiss oscillations as discussed in subsection \ref{subsec:Presence of the uniform vertical electric field}. Simlilarly, when $E_{x}$ is also applied, an effective tilting velocity is introduced and couples with the intrinsic tilting velocity, resulting in valley-polarized Landau level spacing scales and valley-polarized velocity correction factors. These valley-polarized factors break down the spin-valley polarization, releasing the unequal contributions from the spin-up and the spin-down electrons, and finally giving rise to the spin polarization.

\begin{figure}[h]
	\centering
	\subfigbottomskip=2pt
	\subfigcapskip=-5pt
	\subfigure[$\alpha = 0.2, t_{r} = 0.1$]{
		\label{fig:8a}
		\includegraphics[width=0.45\linewidth]{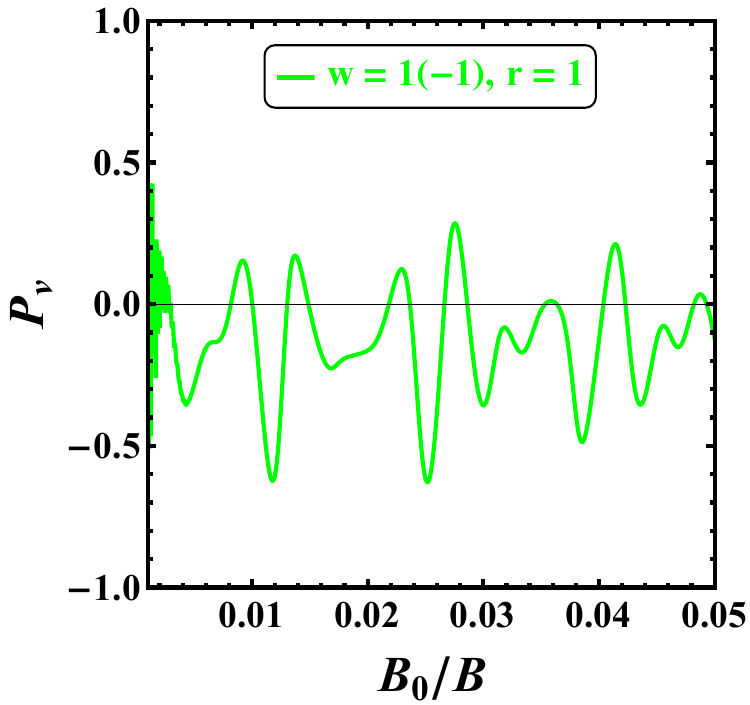}
	}
	\subfigure[$\alpha = 0.2, t_{r} = 0.1$]{
		\label{fig:8b}
		\includegraphics[width=0.45\linewidth]{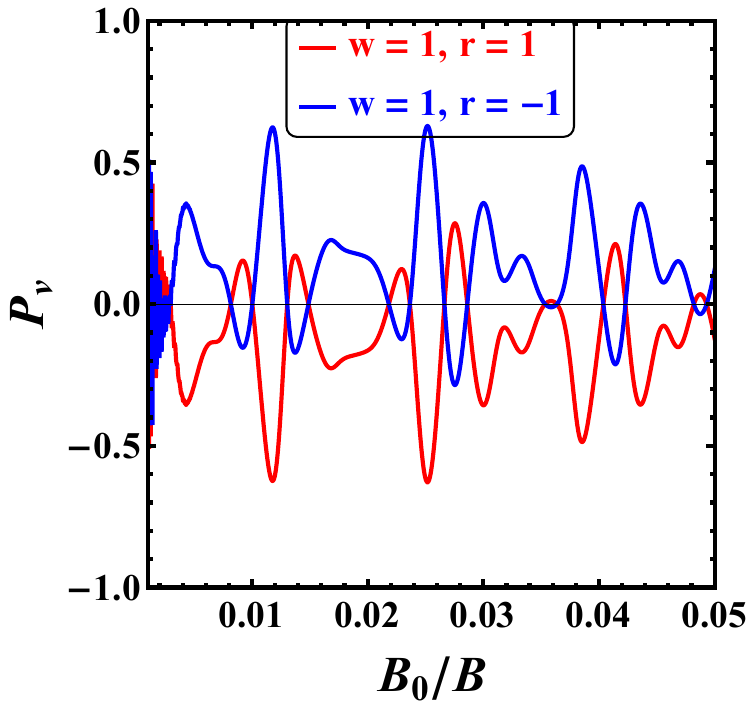}
	}
	\\
	\subfigure[$\alpha = 0.2, t_{r} = 0.1$]{
		\label{fig:8c}
		\includegraphics[width=0.45\linewidth]{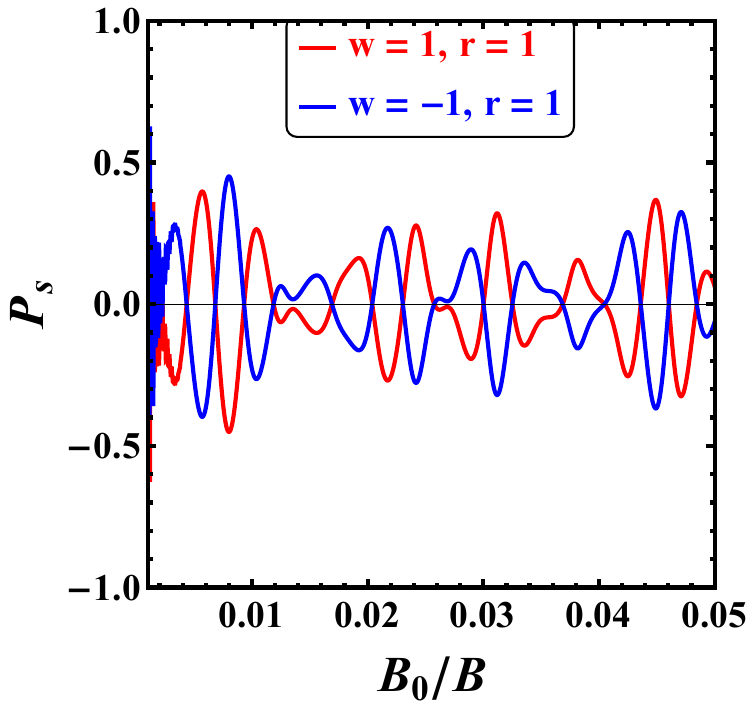}
	}
	\subfigure[$\alpha = 0.2, t_{r} = 0.1$]{
		\label{fig:8d}
		\includegraphics[width=0.45\linewidth]{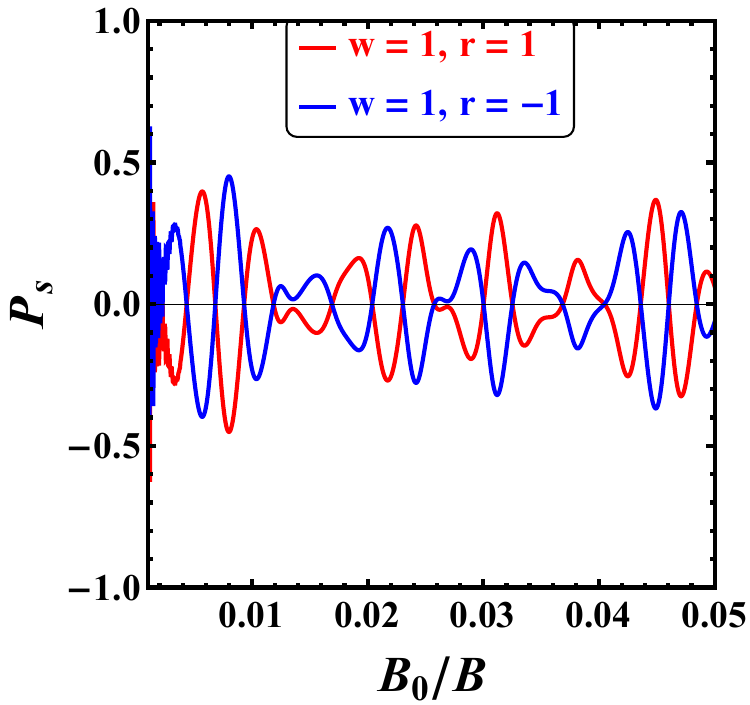}
	}
	\caption{\label{fig:8}The polarization rates versus the inverse magnetic field in the presence of both the uniform vertical and transverse electric fields.}
\end{figure}

Exchanging indices $s,\xi,w,r$ in Eq. (\ref{eq. epsilon}) and Eq. (\ref{eq. Phisxi2}), one finds these relations
\begin{equation}
\label{eq. relation1}
\epsilon_{-s,\xi,\eta,n}(-w) = \epsilon_{s,\xi,\eta,n}(w),
\end{equation}
\begin{equation}
\label{eq. relation2}
\Phi_{\xi}(-w) = \Phi_{\xi}(w),
\end{equation}
\begin{equation}
\label{eq. relation3}
\Phi_{-s}(-w) = \Phi_{s}(w),
\end{equation}
\begin{equation}
\label{eq. relation4}
\epsilon_{-s,-\xi,\eta,n}(-r) = \epsilon_{s,\xi,\eta,n}(r),
\end{equation}
\begin{equation}
\label{eq. relation5}
\begin{split}
&\Phi_{-\xi}(-r) = \Phi_{\uparrow,-\xi}(-r) + \Phi_{\downarrow,-\xi}(-r) \\
&= \Phi_{\downarrow,\xi}(r) + \Phi_{\uparrow,\xi}(r) = \Phi_{\xi}(r),
\end{split}
\end{equation}
\begin{equation}
\label{eq. relation6}
\begin{split}
&\Phi_{-s}(-r) = \Phi_{-s,{\it K}}(-r) + \Phi_{-s,{\it K}^{\prime}}(-r) \\
&= \Phi_{s,{\it K}^{\prime}}(r) + \Phi_{s,{\it K}}(r) = \Phi_{s}(r).
\end{split}
\end{equation}
According to these relations, one can manipulate the polarization of Weiss oscillations in the presence of $E_{z}$ and $E_{x}$, as plotted in Fig.\ \ref{fig:8a}-\ref{fig:8d}. The valley polarization, which originates from the $E_{x}$-introduced valley-polarized Landau level spacing scales and velocity correction factors, can only be switched by flipping $E_{x}$. But the spin polarization, which originates from both the $E_{z}$-introduced spin-valley-polarized effective Fermi energies and the $E_{x}$-introduced valley-polarized factors, can be switched by either flipping $E_{z}$ or $E_{x}$. 

\begin{figure}[h]
\centering
\subfigbottomskip=2pt
\subfigcapskip=-5pt
\subfigure[$w = 1, r = 1$]{
\label{fig:9a}
\includegraphics[width=0.45\linewidth]{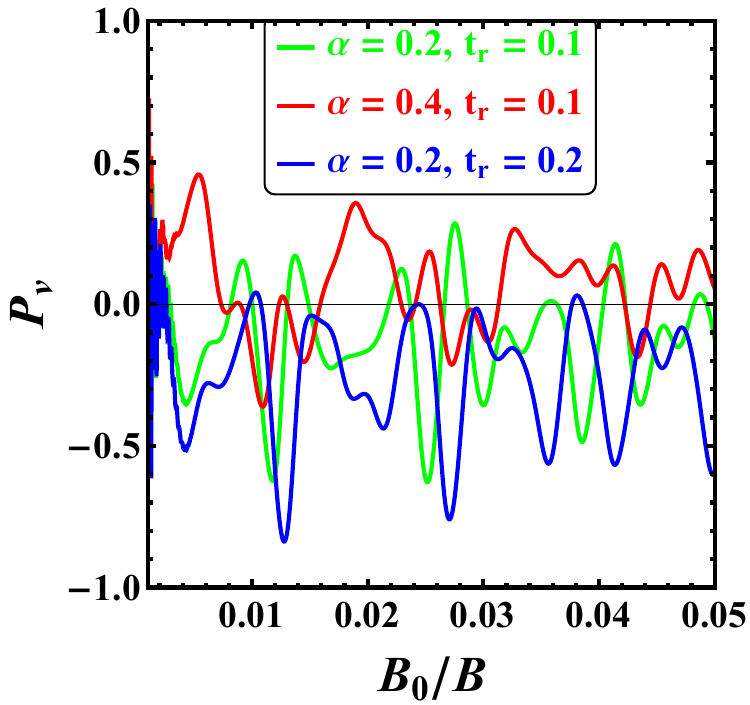}
}
\subfigure[$w = 1, r = 1$]{
\label{fig:9b}
\includegraphics[width=0.45\linewidth]{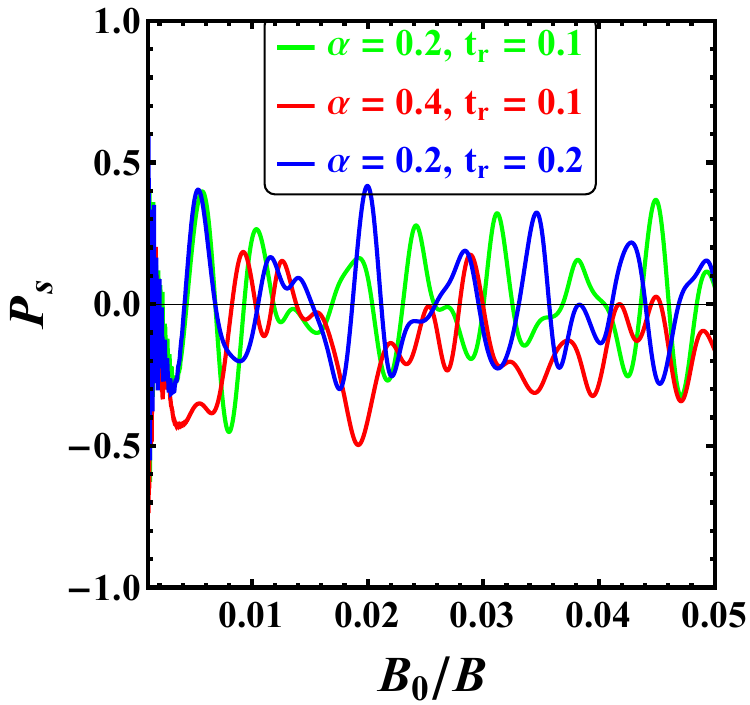}
}
\caption{\label{fig:9}The valley and spin polarization rates versus the inverse magnetic field in the presence of both the uniform vertical and transverse electric fields.}
\end{figure}

In the presence of both the uniform vertical and transverse electric fields, there are also the superposition between the polarization in amplitude and the polarization in period, resulting in a similar valley/spin polarization rate picture as shown in Fig.\ \ref{fig:9a} and Fig.\ \ref{fig:9b}, respectively.

\section{\label{sec:level5}Summary}
\label{sec:5}

We conclude our results from the following three aspects.

Firstly, we have manifested the valley and spin polarization in the Weiss oscillations of the monolayer 1{\it T}$^{\prime}$-$\mathrm{MoS}_{2}$ in the presence of vertical and/or transverse electric fields. It is shown that, when a vertical electric field is applied, the Weiss oscillation curves split into two branches polarized according to the sign of product of the spin index and the valley index (spin-valley-polarized). When a transverse electric field is applied, the Weiss oscillation curves split into two branches polarized according to valley degree of freedom. When both the vertical and transverse electric fields are applied, the Weiss oscillation curves split into four branches polarized according to both spin degree of freedom and valley degree of freedom. 

Secondly, polarizations generated by $E_{z}$ and/or $E_{x}$ can be manipulated. The spin-valley polarization, can be switched by flipping $E_{z}$. The valley polarization, can be switched by flipping $E_{x}$. But the spin polarization, can be switched by either flipping $E_{z}$ or $E_{x}$.

Thirdly, we found that the period polarization in Weiss oscillations originates only from the polarized effective Fermi energies or the polarized Landau level spacing scales. In Weiss oscillations, polarization in amplitude does not imply the presence of polarization in period, whereas polarization in period is invariably accompanied by polarization in amplitude. As long as either the vertical electric field or the transverse electric field exists, both the amplitude and the period of Weiss oscillations become polarized (spin-valley-polarized, or valley-polarized, or both valley-polarized and spin-polarized). However, regardless of which kind of polarization is generated, due to the combination of the amplitude polarization and the period polarization, misalignments of peaks and troughs of oscillation curves from non-equivalent channels have two unique influences on Weiss oscillations. One influence is that although the sign of the polarization observed solely from the perspective of amplitude or period is independent of magnetic field strength, the coupling of these two polarization types yields a polarization rate that can be either positive or negative with respect to different values of the magnetic field strength. The other influence is that although $E_{z}$ and/or $E_{x}$ generate the corresponding polarization here, the strengths of them have little to do with forming a sizable polarization, enabling the appearance of considerable polarization in Weiss oscillations under relatively weak external electric fields.

\begin{acknowledgments}

This work is supported by the National Key R\&D Program of China (Grant No. 2022YFA1403601), the National Natural Science Foundation of China (Grant No. 12504052), the Natural Science Foundation of Jiangsu Province (Grant No. BK20250838), and the Natural Science Foundation of the Jiangsu Higher Education Institutions of China (Grant No. 25KJB140004).
\end{acknowledgments}

\end{document}